# Large transport $J_c$ in Cu-sheathed $Sr_{0.6}K_{0.4}Fe_2As_2$ superconducting tape conductors

**He Lin, Chao Yao, Haitao Zhang, Xianping Zhang, Qianjun Zhang, Chiheng Dong, Dongliang Wang, Yanwei Ma**[*].

Key Laboratory of Applied Superconductivity, Institute of Electrical Engineering, Chinese Academy of Sciences, PO Box 2703, Beijing 100190, China

**Abstract**

Copper sheath is the first choice for manufacturing high-$T_c$ superconducting wires and tapes because of its high electrical and thermal conductivities, low cost and good mechanical properties. However, Cu can easily react with superconducting cores, such as BSCCO, $MgB_2$ and pnictides, during high-temperature sintering process and therefore drastically decrease the transport $J_c$. Here, we report the fabrication of Cu-sheathed $Sr_{1-x}K_xFe_2As_2$ tapes with superior $J_c$ performance using a simple hot pressing method that is capable of eliminating the lengthy high-temperature sintering required by conventional process. We obtained high-quality $Sr_{1-x}K_xFe_2As_2$ tapes with processing at 800°C for 30 minutes and measured high $T_c$ and sharp transition. This rapid fabrication method can effectively thwart the diffusion of Cu into polycrystalline Sr-122 core. As a consequence, we achieved high transport $J_c$ of $3.1 \times 10^4$ A/cm$^2$ in 10 T and $2.7 \times 10^4$ A/cm$^2$ in 14 T at 4.2 K. The in-field $J_c$ performance is by far the highest reported for Cu-sheathed high-$T_c$ conductors. Our results demonstrate the potential of Cu sheath for practical application of pnictide wires and tapes.

[*]Author to whom any correspondence should be addressed.
E-mail: ywma@mail.iee.ac.cn

The discovery of the iron-based superconductors, with a relatively high critical temperature $T_c$, ultrahigh upper critical fields $H_{c2}$ and low anisotropy $\gamma$, has inspired worldwide research efforts[1-5]. Studies on single crystals and thin films reveal that iron-pnictides have large in-field critical current density $J_c$ due to strong intrinsic flux pinning and desirable grain-boundary behavior[6-8]. To explore the potential of using this new material for magnet applications, significant efforts had been focused on developing wires and tapes with in-field high $J_c$. Our research group prepared LaOFeAs$_{1-x}$F$_x$ and SmOFeAs$_{1-x}$F$_x$ superconducting wires using *in-situ* powder-in-tube (PIT) method[5]. A Ta tube or Fe tube with an inner Ti sheath was used to prevent the reaction between the tubes and the superconducting compounds. Zhang *et al.* systematically studied the effect of various sheath materials (Nb, Ta and Fe/Ti) on the microstructure and superconducting properties of SmOFeAs$_{1-x}$F$_x$ wires and reported consistent formation of a thick reaction layer between the core and metal sheath, which hinders the performance of transport $J_c$[9]. Therefore, a key issue is to develop a technology that result in little or no reaction with iron-pnictides. Subsequently, Wang *et al.* reported Ag-sheathed Sr$_{1-x}$K$_x$F$_2$As$_2$ (Sr-122) conductors with transport $J_c$ of 1200 A/cm$^2$ (at 4.2 K and self-field)[10]. The Sr-122/Ag interface was clear and no observable reaction layer, indicating that Ag sheath is benign in proximity to the compound at high sintering temperature. Fe-sheathed Sr-122 and Ba$_{1-x}$K$_x$F$_2$As$_2$ (Ba-122) tapes were also fabricated by *ex-situ* PIT method[11-13]. Through optimization of sintering process and rolling texture, $J_c$ of >10$^4$ A/cm$^2$ in 10 T at 4.2 K was reported for Sr-122 tapes[14]. Recently, Ag-sheathed Ba-122 and Sr-122 conductors have been investigated intensively, and high $J_c$ in exceeding 10$^4$ A/cm$^2$ (Maximum $J_c$=1.2 ×10$^5$ A/cm$^2$) at 4.2 K and 10 T has been reported by applying uniaxial pressing, such as hot pressing (HP) and cold pressing (CP)[15-19]. By far, all pnictide wires and tapes with the high $J_c$-$B$ performance (>10$^4$ A/cm$^2$, at 4.2 K and 10 T) are prepared by using the expensive Ag or magnetic Fe sheath, which is similar to Ag-sheathed BSCCO and Fe-sheathed MgB$_2$ conductors[20-23].

For practical applications of high-$T_c$ superconductors including cuprate, pnictide and MgB$_2$, the copper material is a desirable sheath because of many advantages[24-26].

Firstly, when compared with common Ag and Fe sheath, Cu is a low cost and no-magnetic material. Secondly, Cu sheath has good mechanical properties; the hardness is exactly between Ag and Fe material, which make the coil winding easier in magnet applications. Thirdly, high purity Cu has large residual resistivity ratio (RRR) value, and provides both electromagnetic stabilization against flux jumps and quench protection. However, since the discovery of cuprate superconductors in 1986, no high transport $J_c$-$B$ performance (>$10^4$ A/cm$^2$, at 4.2 K and 10 T) has been reported for Cu-sheathed high-$T_c$ superconductors. Cu is highly reactive to superconducting core at high-temperature sintering [26-29]. The interfacial reaction layer and composition deviation of superconducting phase can lead to $J_c$ degradation. In worse case, no transport $J_c$ can be detected, because the thick reaction layer apparently prevented electric current from flowing from the sheath material to the superconducting core. Therefore, it is considered a grand challenge to develop a process for Cu-sheathed high-$T_c$ superconductors with superior performance. In the present work, we report successfully fabricated Cu-sheathed Sr-122 tapes by an *ex-situ* PIT method. DC susceptibility of Sr-122 precursor powders was measured, and the result is shown in Figure 1. The significant shielding currents appear at about 36.0 K and increase as the temperature decreased, which is similar to that reported for high-quality precursors[19]. During the final heat treatment, we introduce a hot pressing process with combination of short-time sintering (800 ℃/30 min or 700 ℃/60 min) and low external pressure (~20 MPa). This rapid fabrication can effectively avert the formation of reaction layer, and therefore result in high transport $J_c$ of 3.1×$10^4$ A/cm$^2$ at 4.2 K and 10 T.

**Results**

Cu-sheathed Sr-122 tapes were finally hot pressed at 700 ℃ (HP700 tapes) and 800 ℃ (HP800 tapes). Figure 2(a) shows a typical transverse cross-sectional optical image of HP800 tapes. After hot-pressing, the tape thickness of HP800 samples decreased from ~0.40 mm to ~0.29 mm. Figure 2(b) displays a longitudinal cross-sectional optical microstructure of HP800 tapes. A uniform deformation of both superconducting core and Cu sheath along the length can be obviously seen，which is essential for the achievement of high transport $J_c$[30-31]. This uniformity is attributable to the good mechanical properties of Cu sheath.

As shown in Figure 3, the XRD analysis was performed on the planar surfaces of superconducting cores after peeling off Cu sheath. For comparison, the data for randomly orientated precursor powders is also included. The XRD patterns on the surfaces clearly exhibit a $ThCr_2Si_2$-type structure, ensuring that Sr-122 is the main phase for both HP700 and HP800 samples. Using a final short-time hot-pressing process, the formation of non-superconducting reaction layer at the interface seems to be prevented. More importantly, the transport critical current $I_c$ may be measured and obtained in these Sr-122 tapes[27]. However, the impurity peaks are detected on the core surface, especially for HP800 samples. Some Cu reacted with Sr-122 phase, producing SrCuAs and $Cu_{9.5}As_4$ phases. This is consistent with large FWHM (full width at half-maximum) of (002) and (103) peaks for Sr-122 phase. On the other hand, the XRD patterns for the central planar sections of HP tapes after carefully polishing are also exhibited in Figure 3. The diffraction peaks have some differences compared to those of the surfaces. The XRD patterns of central parts exhibit pure Sr-122 phase without detectable impurities. No Cu element can be detected in the central parts by further EDX identification. The peak spreading and symmetry are similar to those of textured Sr-122 tapes[17-19], which have high transport $J_c$-$B$ properties. We quantify the c-axis texture parameter $F$ according to the Lotgering method[32] with $F= (\rho-\rho_0)/ (1-\rho_0)$, where $\rho=\sum I(00l)/ I(hkl)$ and $\rho_0=\sum I_0(00l)/ I_0(hkl)$. $I$ and $I_0$ are the intensities of corresponding XRD peaks measured for the textured and randomly oriented samples, respectively. $F$ values of 0.41 and 0.44 were obtained for HP700 and HP800 tapes, demonstrating that c-axis oriented grains have been achieved in Cu-sheathed tapes. The larger $F$ value in HP800 samples is in agreement with the previous reports confirming that the higher HP temperature, the larger degree of grain alignment[18].

DC susceptibility measurements were conducted on HP700 and HP800 samples. Figure 4(a) depicts two typical groups of the susceptibility curves under a 20 Oe magnetic field parallel to the tape plane. The superconducting transition of HP700 tapes begins at about 33.0 K. It is evident from the zero-field cooled (ZFC) signal that the susceptibility starts to decrease slowly and full shielding is reached at about 15 K. This behavior suggests the presence of inhomogeneity[14, 33]. For HP800 samples, the shielding current occurs at 33.5 K, which may be ascribed to improvement in crystallization. When compared to HP700 samples, the HP800 samples exhibit sharper magnetic transition and reach full shielding at higher temperature (≈20 K). Obviously, enhanced uniformity in superconducting phase has been achieved in HP800 samples[33]. Figure 4(b) shows resistivity versus temperature curves. We measured onset $T_c$ values of 34.6 and 35.1 K for HP700 and HP800 tapes, respectively, which are comparable to Fe-sheathed and Ag-sheathed tapes[34, 17, 19], but slightly smaller than those reported in ref. 18. The impurity of copper compound in present work does not significantly affect the superconducting phase transition. In addition, the resistivity of HP700 and HP800 tapes drops to zero $T_c$ at 32.3 and 33.8 K, respectively. The larger onset $T_c$ and smaller transition width for HP800 samples are consistent with the above magnetic results.

From the viewpoint of practical applications, superconducting wires must be able to carry large transport current density in high magnetic fields. We determined the transport $I_c$ by the standard four-probe method. Figure 5 displays the $J_c$-$B$ properties of HP700 and HP800 tapes at 4.2 K. For HP700 tapes, the $J_c$ values of $3.5 \times 10^4$ A/cm$^2$ and $4.2 \times 10^3$ A/cm$^2$ are obtained in self-field and 10 T, respectively. The striking result is that HP800 tapes show a great enhancement of $J_c$ values in the whole field up to 14 T. Such an improvement can be attributed to improved texture, better homogeneity and crystallization. For HP800 tapes, the $J_c$ data in self-field is not given because the transport $I_c$ is too large (> 450 A), beyond the capability of our existing testing equipment. Excitingly, the transport $J_c$ reaches $3.1 \times 10^4$ A/cm$^2$ at 10 T. To our knowledge, this is by far the highest critical current density under high field ever reported for Cu-sheathed high-$T_c$ superconductors. Importantly, due to its extremely small magnetic field dependence, the transport $J_c$ still maintains a high value of $2.7 \times 10^4$ A/cm$^2$ in 14 T. It is convincible that the Cu-sheathed Sr-122 tapes have a very promising future for use in high-field superconducting magnets.

We conducted SEM/EDX to investigate the influence of hot pressing process on the microstructure of Cu-sheathed Sr-122 tapes. As shown in Figures 6a and 6b, both HP700 and HP800 samples exhibit dense layered structure, which is similar to that of Bi-2223 tapes. HP700 samples have smaller grain size than that of HP800 samples, suggesting that 700 °C hot pressing may not be sufficient. Figure 6c exhibits a typical SEM micrograph of polished cross section of HP800 samples. It is noted that the boundary between Cu sheath and Sr-122 core is clear, further suggesting that there is no apparent reaction layer after hot pressing[13]. The corresponding EDX element mappings of HP800 tapes are presented in Figures 6d-6i. From the Cu mapping, we observe a diffusion of Cu into Sr-122 area, and the diffusion width is approximately 8 μm. This indicates that Cu element interfuses into Sr-122 core and reacts with it during heat treatment. For the elements of Sr-122 phase, Sr, K, Fe and As are detected locally in the core area, disappear almost completely at the border of the metallic area. Comparing with recent Ag-sheathed Sr-122 tapes[18], we conclude that the slightly depression of superconducting properties in this work is mainly due to the diffusion of Cu. At the same time, the diffusion also causes the inhomogeneous distribution of the superconducting elements in Sr-122 area, particularly in the diffusion region. In addition, the EDX mapping of HP700 samples is showed in Figure 6j. For each element, the content has a dramatic change at the border of Sr-122 core. Further analysis reveals that the diffusion width of Cu element is smaller than 3 μm in HP700 samples. Although the sintering time of 60 min is longer than that used for HP800 tapes (30min), the width is much smaller.

The diffusion of Cu and the composition deviation of superconducting phase easily induce severe porosity at the interface, and apparently break the electrical contact between Cu sheath and Sr-122 core[35, 36]. This disadvantageous phenomenon can be avoided by the simple HP method, because it can greatly reduce the pores and cracks by combining the deformation and heat treatment in a single step[17]. As shown in Figure 6a, the Cu sheath and Sr-122 core are tightly connected. As a result, high transport $I_c$ values have been measured in our Cu-sheathed tapes.

For comparison, Cu-sheathed Sr-122 tapes were also sintered without hot-pressing, and the detailed information is exhibited in Table I. The transport $J_c$ values for both HP tapes are much larger than those of corresponding tapes without hot-pressing. For example, the $J_c$ value of HP800 tapes ($3.1 \times 10^4$ A/cm$^2$) is an order of magnitude higher than that of R800 tapes ($3.0\times10^3$ A/cm$^2$), indicating the great $J_c$ enhancement by the hot-pressing method.

**Discussion**

Using copper sheath for superconducting tapes with large transport $J_c$ or $J_e$ is highly desirable for practical applications. By a modified hot-pressing method with combination of final short-time sintering and low external pressure, we successfully prepared Cu-sheathed Sr-122 conductors with large transport current. We demonstrated that the fabricating method developed in our lab can robustly produce high-performance Cu-sheathed superconductors. First, a short-time hot-pressing process can form high-quality Sr-122 phase, which is supported by XRD and resistivity characterizations. For HP800 tapes, the resistivity data demonstrates that the onset $T_c$ is 35.1 K with a transition width of about 1.5 K. Second, this fast fabrication does not give rise to the reaction layer even though the Cu sheath is used. As discussed by above EDX mappings, only a little bit of Cu diffuses into polycrystalline Sr-122 phase. This benefits the superconducting properties, electrical and thermal conductivities of Cu-sheathed tapes. Earlier studies reveal that the thick reaction layer induces the contamination of the superconducting phase to decrease $T_c$, and prevents electric current from flowing from the sheath material to the superconducting core[26-29]. Third, the Cu sheath and Sr-122 core are tightly connected under external pressure, and thus the current path can be enlarged. Meanwhile, the hot pressure can not only considerably increase the core density, but also effectively promote complete reaction of Sr-122 phase, which in return to solve the problem that is low sintering temperature (800 or 700 ℃) and short-time reaction (30 or 60 min) yield poor re-crystallization and ordinary superconducting performance[18, 34]. In summary, the simple hot pressing method ensures high-quality Sr-122 phase and inhibit the formation of reaction layer in Cu-sheathed Sr-122 tapes.

It is fascinating that the largest $J_c$ value of $3.1\times10^4$ A/cm$^2$ in 10 T has been obtained in our best Cu-sheathed tapes. Moreover, the $J_c$ of 122-type pnictides have very weak field dependence in strong fields up to 28 T[37], in accordance with ultrahigh $H_{c2}$ values[5]. Thus, the $J_c$ data above 14 T is given by extrapolating from low fields, as presented in Figure 7. The curve tendency shows that the crossovers with Cu-sheathed NbTi and Nb$_3$Sn wires are around 9.5 and 18.5 T, respectively. This clearly strengthens the position of pnictide conductors as a competitor to the conventional superconductors for high-field applications. On the other hand, Cu-sheathed conductors usually do not need additional stabilization or mechanical reinforcement. Tape conductors with high engineering critical current density ($J_e$) sheathed in comparatively cheap and ductile copper have the strong potential for low specific cost (kA/m) [26, 38]. As showed in Figure 7, a high $J_e$ of about $1.0 \times 10^4$ A/cm$^2$ at 10 T has been achieved in our Cu-sheathed Sr-122 tapes, which is considered as the $J_e$ level desired for practical applications. This achievement is a significant technical breakthrough for the practical applications of Cu-sheathed high-$T_c$ conductors. In the future, if the HP process can be properly adjusted to match the balance between the well re-crystalline reaction and little impurity phase, an even higher $J_e$ can be expected.

**Methods**

**Sample preparation.** We fabricated Cu-sheathed Sr$_{0.6}$K$_{0.4}$Fe$_2$As$_2$ tapes by *ex-situ* PIT method. Sr fillings, K pieces, and Fe and As powder with a ratio of Sr:K:Fe:As = 0.6: 0.5: 2: 2.05 were mixed for 12 hours by ball-milling method. The milled powders were packed into Nb tubes and then sintered at 900 ℃ for 35 h. As prepared Sr-122 superconducting powders were packed into Cu tubes with OD 6 mm and ID 4 mm. These tubes were sealed and then cold worked into tapes (~ 0.4 mm thickness) by swaging, drawing and flat rolling. Finally, hot pressing was performed on the 60 mm long tapes under ~20 MPa at two different sintering processes of 800 ℃/30 min and 700 ℃/60 min. These tapes are defined as HP800 and HP700 tapes, respectively.

**Measurements.** Phase identification of samples was characterized by X-ray diffraction (XRD) analysis with Cu Kα radiation. Magnetization versus temperature curves and resistivity measurements of the superconducting cores were carried out using a PPMS system. The cross sections were polished and then observed by optical images. Microstructure characterization was analyzed using SEM images and EDX scanning. The transport critical current $I_c$ was measured at 4.2 K using short tape samples of 3 cm in length with the standard four-probe method and evaluated by the criterion of 1 μV/cm. The applied fields up to 14 T in $I_c$ measurement were parallel to the tape surface.

# References

1. Kamihara, Y. *et al.* Iron-based layered superconductor La[O$_{1-x}$F$_x$]FeAs (x = 0.05-0.12) with $T_c$ = 26 K. *J. Am. Chem. Soc.* **130**, 3296 (2008).

2. Chen, X. *et al*. Superconductivity at 43 K in SmFeAsO$_{1-x}$F$_x$. *Nature* **453,** 761 (2008).

3. Rotter, M. *et al.* Superconductivity at 38 K in the iron arsenide (Ba$_{1-x}$K$_x$)Fe$_2$As$_2$. *Phys. Rev. Lett.* **101**, 107006 (2008).

4. Putti, M. *et al.* New Fe-based superconductors: properties relevant for applications. *Supercond. Sci. Tech.* **23**, 034003 (2010).

5. Ma, Y. Progress in wire fabrication of iron-based superconductors. *Supercond. Sci. Technol.* **25,** 113001 (2012).

6. Yang, H. *et al.* Fishtail effect and the vortex phase diagram of single crystal Ba$_{0.6}$K$_{0.4}$Fe$_2$As$_2$. *Appl. Phys. Lett.* **93**, 142506 (2008).

7. Wang, X. *et al.* Very strong intrinsic flux pinning and vortex avalanches in (Ba, K)Fe$_2$As$_2$ superconducting single crystals. *Phys. Rev. B* **82**, 024525 (2010).

8. Katase, T. *et al.* Advantageous grain boundaries in iron pnictide Superconductors. *Nat. Commun.* **2**, 409 (2011).

9. Zhang, X. *et al*. Effect of sheath materials on the microstructure and superconducting properties of SmO$_{0.7}$F$_{0.3}$FeAs wires. *Physica C* **470,** 104 (2010).

10. Wang, L. *et al.* Large transport critical currents of powder-in-tube Sr$_{0.6}$K$_{0.4}$Fe$_2$As$_2$/Ag superconducting wires and tapes. *Physica C* **470**, 183 (2010).

11. Wang, L. *et al.* Textured Sr$_{1-x}$K$_x$Fe$_2$As$_2$ superconducting tapes with high critical current density. *Physica C* **471,** 1689 (2011).

12. Zhang, X. *et al*. Mechanism of enhancement of superconducting properties in a Ba$_{1-x}$K$_x$Fe$_2$As$_2$ superconductor by Pb addition. *Supercond. Sci. Technol.* **25,** 084024 (2012).

13. Yao, C. *et al.* Improved transport critical current in Ag and Pb co-doped Ba$_x$K$_{1-x}$Fe$_2$As$_2$ superconducting tapes. *Supercond. Sci. Technol.* **25,** 035020 (2012).

14. Gao, Z. *et al.* High critical current density and low anisotropy in textured Sr$_{1-x}$K$_x$Fe$_2$As$_2$ tapes for high field applications. *Sci. Rep.* **2,** 998 (2012).

15. Togano, K. *et al.* Enhancement in transport critical current density of *ex situ* PIT

Ag/(Ba, K)Fe$_2$As$_2$ tapes achieved by applying a combined process of flat rolling and uniaxial pressing. *Supercond. Sci. Technol.* **26,** 115007 (2013).

16. Gao, Z. *et al.* Achievement of practical level critical current densities in Ba$_{1-x}$K$_x$Fe$_2$As$_2$/Ag tapes by conventional cold mechanical deformation. *Sci. Rep.* **4**, 4065 (2014).

17. Lin, H. *et al.* Strongly enhanced current densities in Sr$_{0.6}$K$_{0.4}$Fe$_2$As$_2$+Sn superconducting tapes. *Sci. Rep.* **4**, 4465 (2014).

18. Lin, H. *et al.* Hot pressing to enhance the transport $J_c$ of Sr$_{0.6}$K$_{0.4}$Fe$_2$As$_2$ superconducting tapes. *Sci. Rep.* **4,** 6944 (2014).

19. Zhang, X. *et al.* Realization of practical level current densities in Sr$_{0.6}$K$_{0.4}$Fe$_2$As$_2$ tape conductors for high-field applications. *Appl. Phys. Lett.* **104**, 202601 (2014).

20. Yuan, Y. *et al.* Significantly enhanced critical current density in Ag-sheathed (Bi, Pb)$_2$Sr$_2$Ca$_2$Cu$_3$O$_x$ composite conductors prepared by overpressure processing in final heat treatment. *Appl. Phys. Lett.* **84**, 2127 (2004).

21. Larbalestier, D. C. *et al.* Isotropic round-wire multifilament cuprate superconductor for generation of magnetic fields above 30 T. *Nat. Mater.* **13**, 375 (2014).

22. Hossain, M. S. A. *et al.* The enhanced $J_c$ and $B_{irr}$ of *in situ* MgB$_2$ wires and tapes alloyed with C$_4$H$_6$O$_5$ (malic acid) after cold high pressure densification. *Supercond. Sci. Technol.* **22,** 095004 (2009).

23. Herrmann, M. *et al.* Touching the properties of NbTi by carbon doped tapes with mechanically alloyed MgB$_2$. *Appl. Phys. Lett.* **91** 082507 (2007).

24. Wilson, M. N. *Superconducting Magnets* (Clarendon, 1987).

25. Cooley, L. D. *et al.* Costs of high-field superconducting strands for particle accelerator magnets. *Supercond. Sci. Technol.* **18,** R51 (2005).

26. Wozniak, M. *et al.* Study of short duration heat treatments of an *in situ* copper-sheathed MgB$_2$ wire. *Supercond. Sci. Technol.* **23,** 105009 (2010).

27. Ding, Y. *et al.* Effects of cold high pressure densification on Cu sheathed Ba$_{0.6}$K$_{0.4}$Fe$_2$As$_2$ superconducting wire. *Physica C* **483,** 13 (2012).

28. Nakamura, Y. *et al.* Effect of Bi2223 addition in precursor on the formation and $J_c$ property of Bi2223 tapes sheathed with the Ag–Cu alloy. *Supercond. Sci. Technol.*

**21,** 035001 (2008).

29. Pachla, W. *et al.* Structural characterization of multifilament heat treated *ex situ* MgB$_2$ superconducting wires with Cu and Fe sheaths. *Supercond. Sci. Technol.* **17,** 1289 (2004).

30. Han, Z. *et al.* The mechanical deformation of superconducting BiSrCaCuO/Ag composites. *Supercond. Sci. Technol.* **10,** 371 (1997).

31. Osamura, K. *et al.* Work instability and its influence on the critical current density of silver sheathed Bi2223 tapes. *Supercond. Sci. Technol.* **5,** 1 (1992).

32. Lotgering, F. Topotactical reactions with ferrimagnetic oxides having hexagonal crystal structures-I. *J. Inorg. Nucl. Chem.* **9**, 113 (1959).

33. Palenzona, A. *et al.* A new approach for improving global critical current density in Fe(Se$_{0.5}$Te$_{0.5}$) polycrystalline materials. *Supercond. Sci. Technol.* **25**, 115018 (2012).

34. Lin, H. *et al.* Effects of heating condition and Sn addition on the microstructure and superconducting properties of Sr$_{0.6}$K$_{0.4}$Fe$_2$As$_2$ tapes. *Physica C* **495**, 48 (2013).

35. Husek, I. *et al.* Microhardness as a tool for the filament density and metal sheath analysis in MgB$_2$/Fe/(Cu) wires. *Supercond. Sci. Technol.* **17,** 971 (2004).

36. Flukiger, R. *et al.* Superconducting properties of MgB$_2$ tapes and wires. *Physica C* **385**, 286 (2003).

37. Gao, Z. *et al.* High transport Jc in magnetic fields up to 28T of stainless steel/Ag double sheathed Ba122 tapes fabricated by scalable rolling process. *Supercond. Sci. Technol.* **28,** 012001 (2015).

38. Weiss J. D. et al. High intergrain critical current density in fine-grain (Ba$_{0.6}$K$_{0.4}$)Fe$_2$As$_2$ wires and bulks. *Nat. Mater.* **11**, 682 (2012).

**Acknowledge**

The authors thank Professors S. Awaji and K. Watanabe for the transport $I_c$ measurements. We acknowledge suggestions and discussions with Dr. B. Ma at ANL. This work is partially supported by the National '973' Program (grant No. 2011CBA00105), the National Natural Science Foundation of China (grant Nos. 51172230, 51320105015 and 51202243) and the Beijing Municipal Science and Technology Commission (grant No. Z141100004214002).

**Author contributions**

Y. W. M. directed the research. H. L. fabricated the tape samples. H. L. carried out XRD, magnetization, resistivity, and microstructure characterizations. X. P. Z. and D. L. W. did the high-field $I_c$ measurement. Q. J. Z. and C. H. D helped with the tape preparation. C.Y. and H. T. Z. contributed to manuscript preparation. H. L. and Y. W. M. wrote the manuscript. All the authors contributed to discussion on the results for the manuscript.

**Additional information**

Competing financial interests: The authors declare no competing financial interests.

Table I. The onset $T_c$ and transport $J_c$ for Cu-sheathed Sr-122 tapes with and without hot pressing.

| Sample name | Fabrication conditions | Onset $T_c$ (K) | $J_c$ (kA/cm$^2$) (4.2K, 10T) |
|---|---|---|---|
| R700 | 700ºC/1h | 34.3 | 1.3 |
| R800 | 800ºC/0.5h | 34.5 | 3.0 |
| HP700 | 700ºC/1h/20MPa | 34.6 | 4.2 |
| HP800 | 800ºC/0.5h/20MPa | 35.1 | 31.0 |

**Captions**

Figure 1| Magnetization versus temperature curves of the precursor powders.

Figure 2| Optical images of Cu-sheathed Sr-122 tapes: (a) the transverse cross section of HP800 tapes; (b) the longitudinal cross section of HP800 tapes.

Figure 3| XRD patterns for the planar surfaces and central sections of the superconducting cores of both HP700 and HP800 samples. As a reference, the data for randomly orientated precursor powders is also included. The peaks of $Sr_{1-x}K_xFe_2As_2$ phase are indexed, while the peaks of impurity phases are also marked.

Figure 4| (a) Temperature dependence of the DC magnetic susceptibility curves of HP700 and HP800 samples. (b) Resistivity versus temperature curves of HP700 and HP800 samples; Inset showing the enlarged part near the superconducting transition. All data were obtained after peeling off the Cu sheath.

Figure 5| Magnetic field dependence of the transport $J_c$ at 4.2 K for both HP700 and HP800 tapes. The applied fields up to 14 T were parallel to the tape surface.

Figure 6| SEM microstructures of Sr-122 tapes: (a) HP700 and (b) HP800 samples. (c) SEM image showing the interface between Cu sheath and Sr-122 core for HP800 tapes. (d) The corresponding EDX mapping image for HP800 samples; (e-i) area mappings of Cu, Sr, K, Fe and As element, respectively. (j) The EDX mapping image for HP700 samples.

Figure 7| Magnetic field dependence of the transport critical current density $J_c$ and the engineering current density $J_e$ for our best Sr122/Cu tapes. The $J_c$-$B$ curves of PIT processed NbTi and $Nb_3Sn$ wires are also shown for comparison.

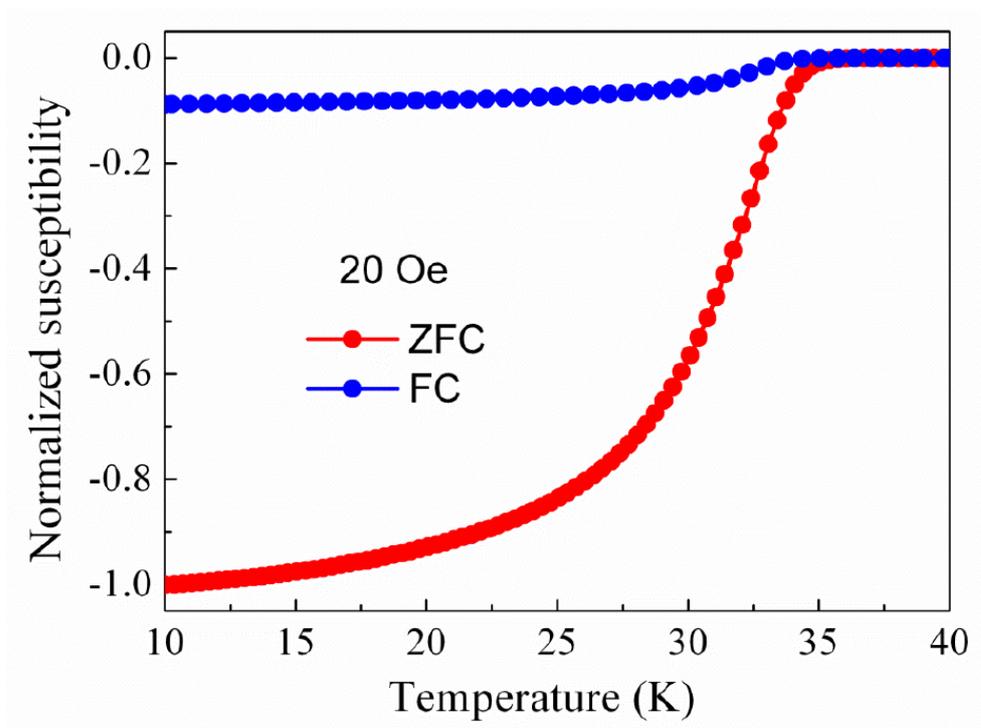

Figure-1 Lin et al.

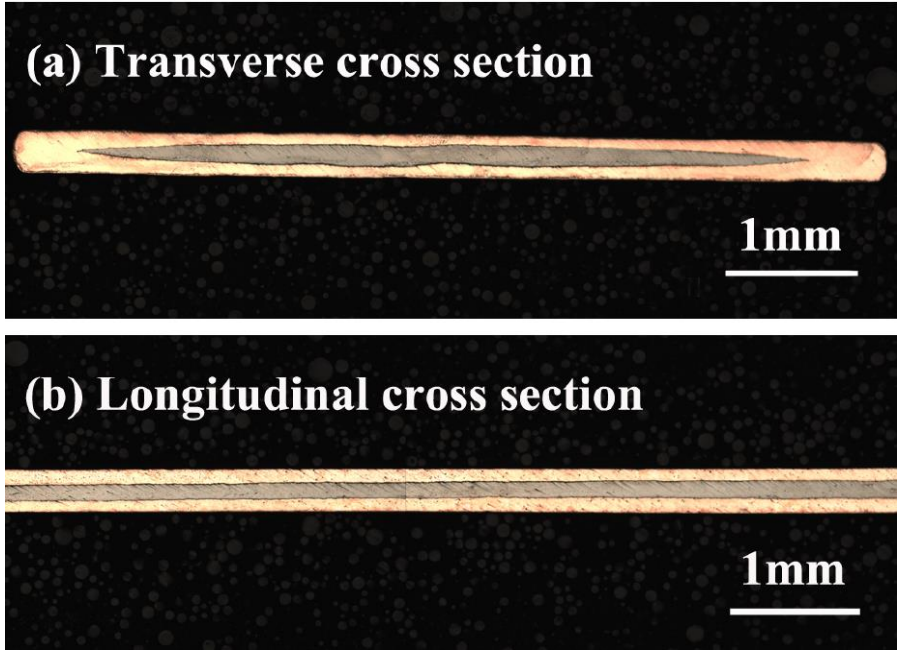

Figure-2 Lin et al.

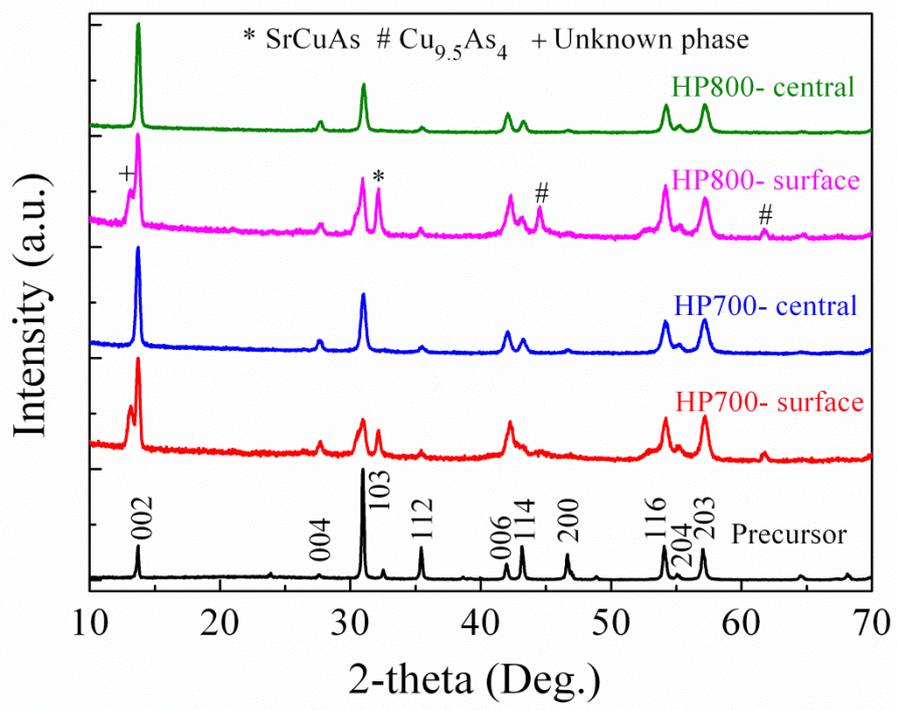

Figure-3 Lin et al.

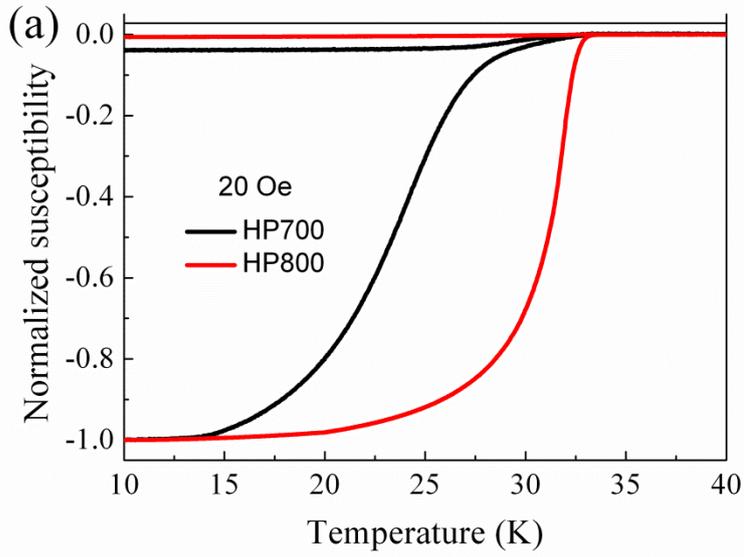

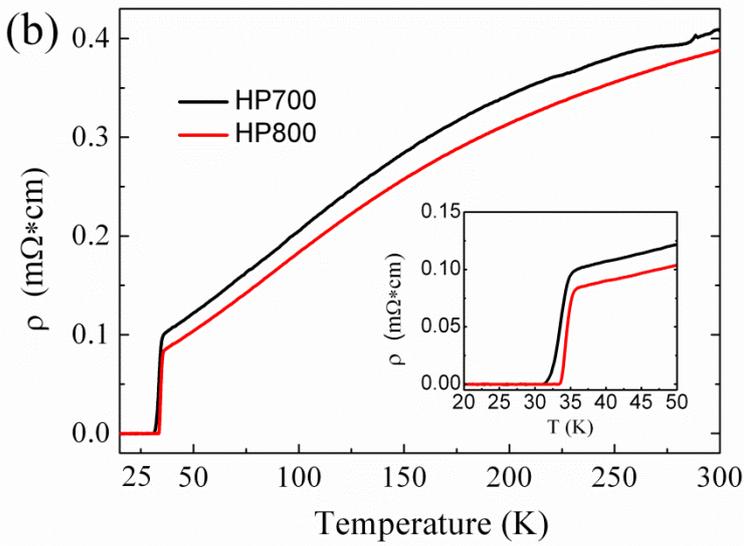

Figure-4 Lin et al.

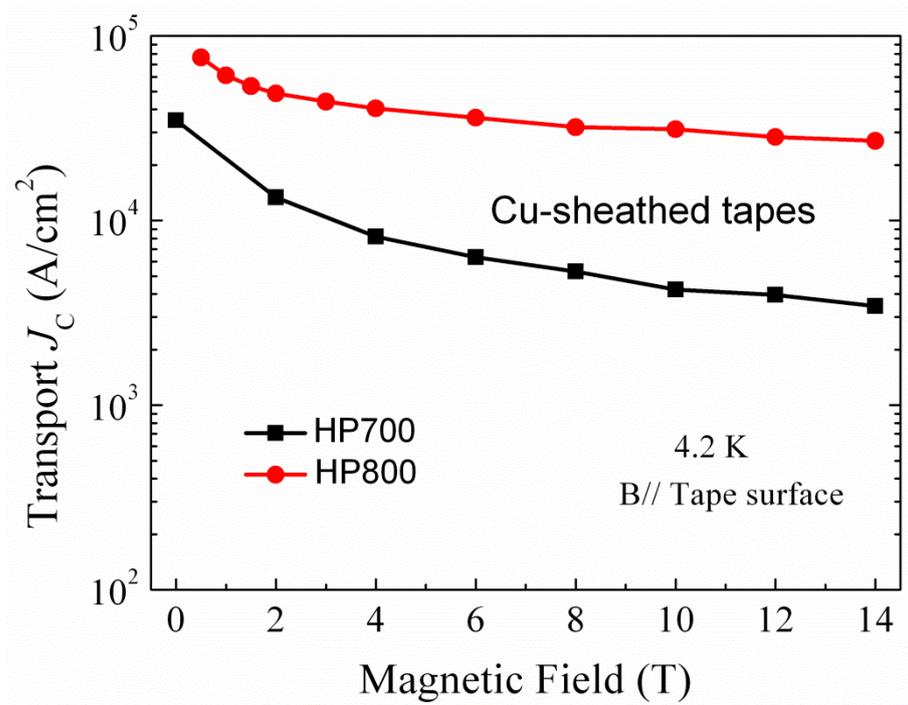

Figure-5 Lin et al.

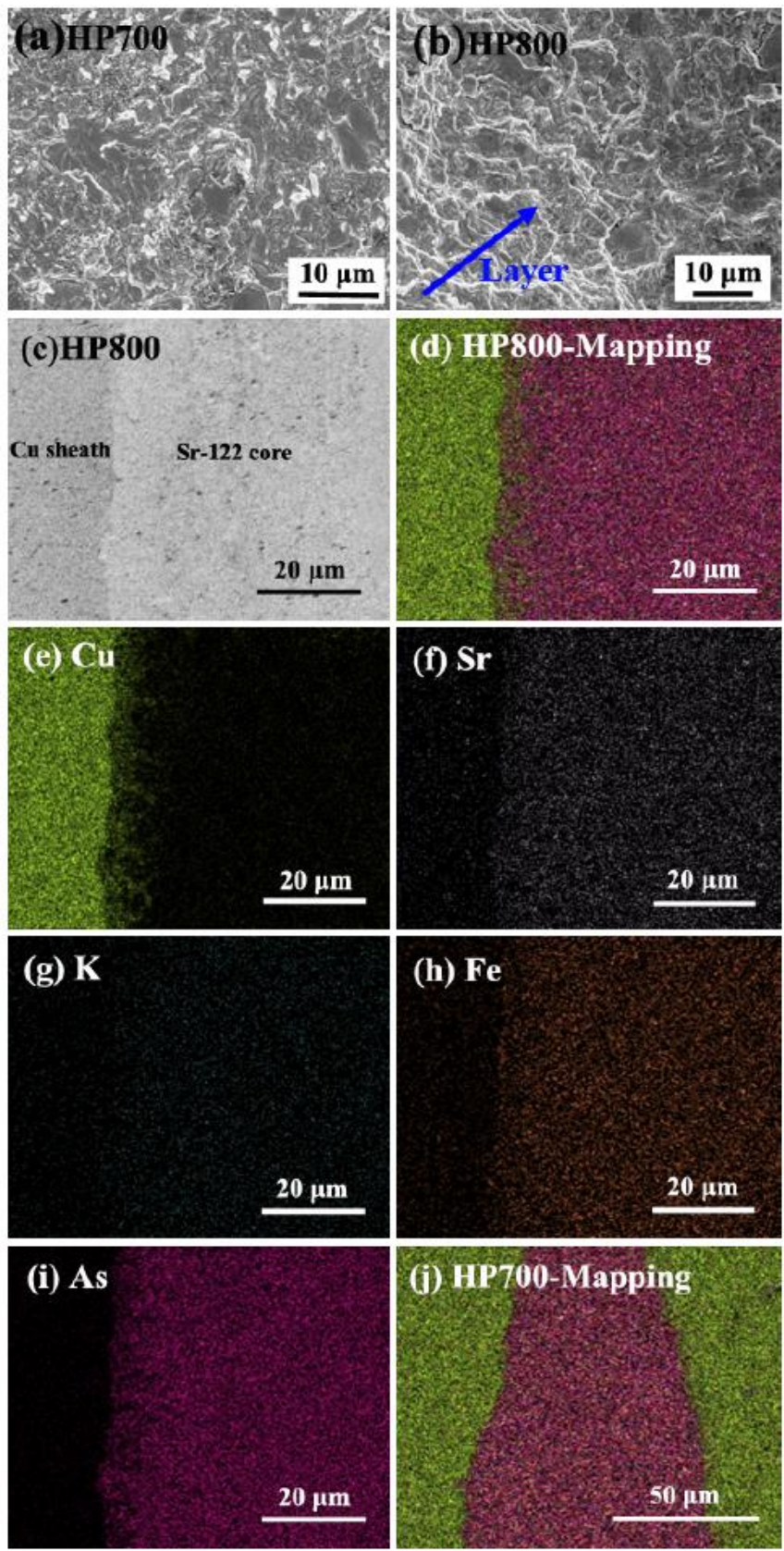

Figure-6 Lin et al.

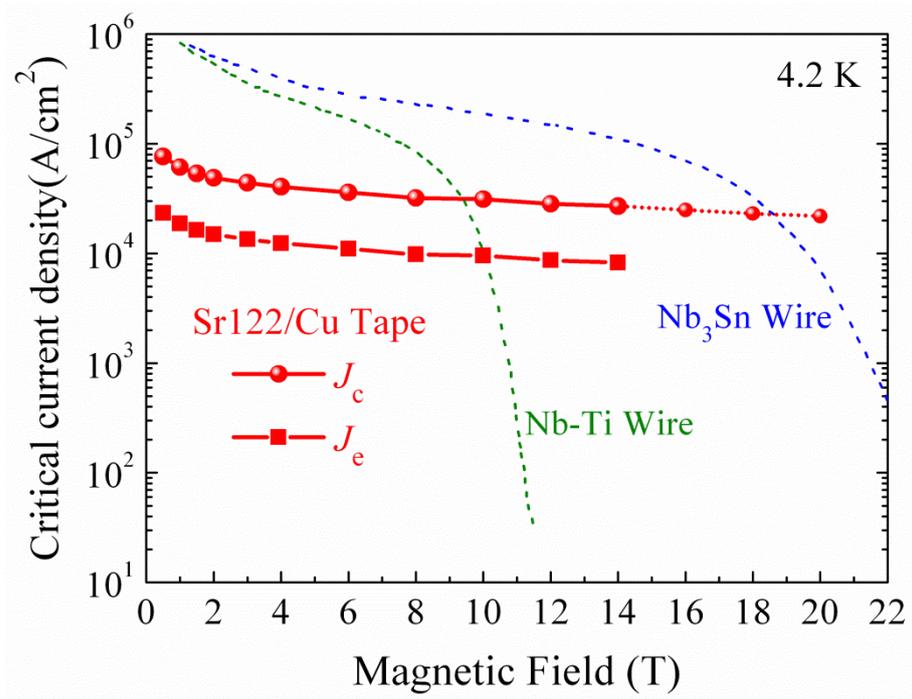

Figure-7 Lin et al.